\documentstyle[aps,prl,psfig]{revtex}
\begin{document}
\title{ Fragilities of Liquids Predicted from the Random First Order Transition Theory of Glasses\\}
\author{Xiaoyu Xia and Peter G. Wolynes \\
Departments of Physics and Chemistry\\ 
University of Illinois, Urbana, IL, 61801\\} 
\date{December 16, 1999}
\maketitle

\begin{abstract}
{A microscopically motivated theory of glassy dynamics based on an underlying random first order transition is developed to explain the magnitude of free energy barriers
for glassy relaxation.  A variety of empirical correlations embodied in the
concept of liquid ``fragility" are shown to be quantitatively explained by such a model.  The near universality of a Lindemann ratio characterizing the maximal amplitude of thermal vibrations within an amorphous minimum explains the variation of fragility with a liquid's configurational heat capacity density.  
Furthermore the numerical prefactor of this correlation is well approximated 
by the microscopic calculation.  The size of heterogeneous reconfiguring
regions in a viscous liquid is inferred and the correlation of nonexponentiality
of relaxation with fragility is qualitatively explained.  Thus the wide variety of kinetic behavior in liquids of quite disparate chemical nature reflects quantitative rather than qualitative differences in their energy landscapes.
}
\end{abstract}
\vspace{1cm}
It is believed all classical fluids could form glasses if cooled 
sufficiently fast so as to avoid crystallization.  Central to glass formation is a 
dramatic slowing of molecular motions on cooling the liquid.  The existence of a 
universal description of glass transitions is suggested by empirical 
observations connecting deviations from the Arrhenius law for the slowing of 
rates, nonexponential relaxations in the super cooled liquid state and the 
behavior of thermodynamic properties on cooling(1).  Quantitative differences in 
behavior of different substances sometimes obscure this universality.  This has led to a classification of liquids into ``fragile" 
ones like o-terphenyl, having the most dramatic deviations from the 
Arrhenius law, and into ``strong" ones like pure SiO$_2$ where the Arrhenius 
equation works well(1).  In this paper, we 
show how the fragile versus strong behavior of liquids can be understood within 
a microscopically motivated theory based on the idea that glassy dynamics is 
caused by an underlying thermodynamic, ideal ``random first order" transition(2-7).

The notion that a random first order transition lies at 
the heart of glass formation received its early theoretical support from the remarkable 
confluence of approximate microscopic theories of the liquid glass transition(8-10) and 
the behavior of a large class of exactly solvable statistical 
mechanical models of spin glasses with quenched disorder(11).  Two closely connected theories 
of the liquid glass transition suggest features similar to first order transitions.  One of these, the so-called mode-mode coupling theory(8,12), focuses on the feedback between the slow fluctuations of 
fluid density in a molecule's environment on the motion of that molecule.  
This theory predicts a sharp transition in the dynamics as well as a 
characteristic behavior of the time correlation functions near the 
predicted transition.  M\"{o}ssbauer effect(13) and neutron scattering(14) are roughly consistent with these precursor phenomena.  At temperatures below the transition, mode coupling theory predicts the freezing of the liquid's configuration near to a given random configuration; i.e., 
there is broken ergodicity.  Another approach to the glass 
transition directly addresses broken ergodicity by investigating the stability 
of a frozen density wave using either self-consistent phonon theory(9) or 
the density functional theory of liquids, applying them to aperiodic structures 
(10,15).  The mode coupling, self-consistent phonon and density functional 
approaches all predict that there is a Lindemann criterion for the stability of 
an aperiodic density wave: just as for a periodic crystalline solid, 
thermal vibrations cannot yield a root mean square displacement 
of particles from their fiducial location exceeding roughly one tenth of the interparticle spacing.  The precise value of the Lindemann ratio only weakly depends on the detailed intermolecular forces.  The predicted Lindemann ratio corresponds well to the experimentally measured magnitude of the 
intermediate time plateau in the structure function measured by neutrons(14).  A finite Lindemann ratio would be consistent with a first 
order phase transition, but glass transitions in the laboratory do not 
show a latent heat as ordinary first order transitions do. This lack of latent 
heat is explained by the existence of the large number of aperiodic 
structures that may be frozen in at a glass transition in contrast to the unique 
periodic structure formed in ordinary crystallization.  Many exactly 
solvable models of disordered magnetic systems have been shown to exhibit freezing into many structures(11,16,17).  The major class of these also show a first order 
jump in a locally defined order parameter without any latent heat.   This defines what has been called a ``random first order" transition. Unlike Ising spin glasses, these models possess no symmetry between local states but have long range, 
quenched random interactions.  Such systems 
include Potts spin glasses(11), $p$-spin glasses(17), and the 
elegantly solved Random Energy Model(16).  There are further parallels between these 
systems and the phenomenology of glass forming liquids, most notably both glass forming liquids and these models exhibit a Kauzmann entropy crisis, i.e., the configurational entropy vanishes at a finite temperature above absolute zero(18).  This crisis would define an underlying ideal glass transition.  Whether the crisis for liquids would be avoided in some way at lower temperature than measurements have been made is controversial and is of limited relevance to describing the observed behavior using the analogy.  In the 
exactly solvable statistical mechanical models, a dynamic transition occurs at a high temperature T$_A$ 
coincident with mode coupling and stability analyses, but the thermodynamic 
transition does not occur until at a lower temperature, T$_K$ the configurational entropy of different frozen solutions vanishes(4).  The idea then is that the glassy dynamics in the measured temperature range is governed by the approach to an ideal glass transition described like that shown in the exactly solved models.  There are two seeming differences between the exactly solved models and the situation for the liquid-glass transition.  First, in 
liquids there is no quenched randomness; it must be 
self-generated. Second, while the models have infinite range forces, interactions in liquids are of finite range.  The absence of quenched randomness has been addressed by exhibiting several mean 
field models without quenched randomness that do generate randomness internally(19-21).  Also the formal statistical mechanical tools used for quenched random Hamiltonians, e.g., the replica technique, have been shown to be 
applicable to atomic fluid systems with self-generated randomness(7).  
Furthermore, computer simulations of fluid glass transitions show replica symmetry breaking like a random first order 
Transition(22).  The consequences of finite range interactions are more important.  The finite range causes the dynamic transition at T$_A$, like a spinodal of an ordinary first order transition to be smeared out.  It becomes a crossover to activated dynamics.  Below T$_A$, motions in the 
finite 
range system can still occur that involve the rearrangement of large regions of the 
liquid.  The transition to such collective activated events in liquids has been confirmed in simulations(23,24).  The events are driven by the configurational entropy.  For 
finite range systems approaching a random first order transition, an ``entropic droplet" 
scaling argument for the activation barriers naturally explains the 
non-Arrhenius transport behavior and leads to the Vogel-Fulcher law(4,5).  The idea that configurational entropy is needed for motions in glasses predates the random first order transition theory and was described by Adam and Gibbs(25).  The older argument is really quite different from the random first order transition theory, since it provides no explanation for how a rearranging unit's activation energy is related to the microscopic forces.  Here we show how the near universality of the Lindemann ratio explains the 
connection between barrier heights and thermodynamics for liquids of varying 
fragility.

The naive density functional approach used to obtain the Lindemann 
criterion for vitrification allows an estimate for the free energy of dynamic 
rearrangements.  The density functional(10,26) assesses the cost 
of forming any density wave by breaking the free energy into an entropic 
localization penalty and an interaction term. 
\begin{equation}\label{eq:fefn}
F=\int f(\rho({\bf r}))d^3{\bf r}=k_BT\int d^3{\bf r}\rho({\bf r})[\ln \rho ({\bf 
r})-1]+\int\!\!\int d^3{\bf r}d^3{\bf r'} (\rho({\bf r})-\rho_0)c({\bf r}-{\bf 
r'})(\rho({\bf r'})-\rho_0),
\end{equation}
where $\rho_0$ is the mean density.
The localization cost is the same as for a perfect gas while the 
interaction term involves the direct correlation function of the liquid, a renormalized form of the bare interaction potential.  The 
direct correlation function is determined by the condition that the functional gives small 
fluctuations in density reproducing the static liquid structure factor.  Higher order terms in the density can also be included.  In 
the frozen aperiodic state the density wave is decomposed into a sum of Gaussians 
centered around random lattice sites, $\rho({\bf r})=\sum_{i}  
(\frac{\pi}{\alpha})^{3/2}\exp (-\alpha({\bf r}-{\bf r}_i)^2)$, where $\alpha$ 
represents the effective local spring constant that determines the rms displacement from the fiducial lattice site.  The localization sites are $\{{\bf r}_i\}$.  For large 
$\alpha$, 
the densities around different sites overlap weakly giving
\begin{equation}\label{eq:fef}
\frac{F}{N} =k_BT [\frac{3}{2} \ln (\frac{\alpha r_0^2}{\pi} )-\frac{5}{2}]+\frac{1}{N} \int\!\!\int d^3{\bf r}d^3{\bf  
r'} 
(\rho({\bf r})-\rho_0)c({\bf r}-{\bf r'})(\rho ({\bf r'})-\rho_0),
\end{equation}
where $N$ is the total number of particles and $r_0$ is the mean lattice spacing.  We can take $\rho _0r_0^3=1$.
For small $\alpha$, $F/N$ reduces to the perfect gas value.

A similar free energy expression is obtained from self-contained phonon theory where the direct correlation function is replaced by the 
Mayer function $f=e^{-\beta u(r)}-1$ for hard potentials(10) or by the potential itself(7).  The free energy varies 
with the particular arrangement of sites $\{{\bf r}_i\}$, but assuming 
$\alpha$ 
is the constant, the mean free energy of aperiodic structures is plotted in Figure 1(a).  The 
lowest value of $\alpha$ for which a secondary minimum occurs is given by 
the 
Lindemann value $\alpha _L$.  This minimum representing the frozen wave is 
higher 
in free energy than the $\alpha =0$ fluid phase.  For the exactly solvable random first order transitions the excess free energy of the frozen solution is known to equal the configurational entropy of 
possible mean field solutions, $TS_c$.  For the fluid system in addition to the $\alpha=0$ and $\alpha \approx \alpha _L$ 
uniform stationary solutions of the variational equation $\delta F=0$, there are saddle points representing droplet 
configurations in which a region of low $\alpha \approx 0$ forms in the 
midst 
of a given large $\alpha$ solution (See Figure 1(b)).  This saddle point is a transition state for reconfiguring the frozen density 
wave.  Within the melted region there is a multiplicity of states corresponding to other aperiodic arrangements of the atoms.  Much below T$_A$, the interface should be quite sharp, that is, in 
a 
single atomic layer $\alpha$ changes from a value near $\alpha_L$ to near zero.  Close to T$_A$ 
the 
transition should be smoother with $\alpha$ slowly varying over many atomic 
layers.  In both cases, there will arise a surface tension $\sigma$
reflecting the deviation of $\alpha$ in the layers with $\alpha$ different from 
the bulk free energy minima values.  The density functional 
expression for the droplet free energy then is given as a function of the radius 
of the droplet much as in conventional nucleation,
\begin{equation}\label{eq:mffe}
F(r)=-\frac{4}{3}\pi Ts_cr^3+4 \pi \sigma r^2.
\end{equation}
Here $s_c$ is the configurational entropy density.

The maximum of $F(r)$ gives a reconfiguration barrier $\Delta F^{\ddagger}=\frac{16}{3} \pi \sigma^3/(Ts_c)^2$.  A detailed calculation of this barrier for a specific glassy system, the random heteropolymer has been given by Takada and Wolynes(27).  This naive droplet result(4,28) differs from the 
Adam Gibbs suggestion $\Delta F^{\ddagger}=s_c^{\ast}\Delta\mu/s_c$(25), where $\Delta\mu$ is a bulk ``activation energy" per particle and $s_c^{\ast}$ is the ``critical configurational entropy" taken to be usually $k_B\ln 2$.  The AG formula is not the result of a self-contained microscopic calculation but assumes the free energy cost of dynamically reconfiguring a region is independently given from the free energies that
determine the low energy structures themselves.  There is no apparent reason to assume $\Delta \mu$ a constant for different substances. On the other hand, the modern random first order transition 
theory does suggest universality for $\sigma$ based on the universality of the 
Lindemann ratio $\alpha _L^{-1/2}/r_0$.  We see this in the following way: assuming a sharp interface between the localized and delocalized regions, the energy associated 
with 
the interface should be one-half of the interaction part of the free energy in 
the bulk stable phase.  Therefore, $\sigma = \frac{T}{2}r_0 [\frac{3}{2} nk_B\ln (\frac{\alpha r_0^2}{\pi e}) -s_c(T)]$, where $n$ is the density of particles.
Since the localization part of the free energy depends only logarithmically on 
$\alpha$, we can replace $\alpha$ by its minimum value $\alpha_L$ which it achieves at T$_A$.  Near T$_K$, on the 
other hand, we can neglect the configurational entropy part of the expression.  For 
temperatures between T$_A$ and T$_K$, the errors of making these two approximations largely cancel.  This gives $\sigma = \frac{3}{4}nr_0k_BT \ln(\frac{\alpha_Lr_0^2}{\pi e})= \sigma _0$ as an approximation for temperatures much below 
T$_A$.  The universality of the Lindemann ratio $\alpha _L^{-1/2}/r_0$ means $\sigma/nr_0k_BT$ is nearly universal and therefore that $\Delta F^{\ddagger}$ increases 
more 
rapidly with cooling for substances with a large 
configurational heat capacity. This explains the empirical correlation that strong liquids with nearly Arrhenius rate slowing have small excess 
heat capacities contrasting with fragile liquids having large excess heat 
capacity with dramatically non-Arrhenius slowing.  Near T$_A$ the interface 
broadens and the sharp interface approximation breaks down.  
A gradient expansion of the free energy as a function of $\alpha$ yields a surface energy vanishing near T$_A$.  The universal value of $\sigma _0$ is only approximate.  For given substance, the remaining temperature dependence of $\sigma$ from the broadening of the interface implies that the apparent fragility of liquids measured at high temperature should be larger than that 
measured at low temperature, as noted by Angell in his detailed survey of viscosity data(29).  Similarly we note that $\sigma$ depends on the density and therefore the pressure.  Thus although a kinetic glass transition defined by a specific numerical barrier height or fiducial relaxation time will be largely a function of the configurational entropy density there will be another explicit but weak thermodynamic dependence on pressure too.  Consistent with Nieuwenhuizen's recent analysis of the dynamic effects on glass transitions caused by pressure and temperature change(30), this could explain the mild deviation of the Prigogine-deFay ratio from 1.

While the simple density functional calculation explains qualitatively the 
fragility/heat capacity density correlation, viscosity data are more consistent with an
$s_c^{-1}$ scaling for the free energy of activation (like that suggested by Adam and Gibbs(25) ) rather than the $s_c^{-2}$ behavior predicted from the simple 
density functional theory.  The scaling theory of the entropic droplet formulation already accounts for this observation(5).  The modification comes from the complexity of the interface between aperiodic crystalline minima(5).  
Correct scaling near T$_K$ is restored by the wetting of droplets 
corresponding with one particular density wave, by a surface coating 
corresponding to a different aperiodic arrangement.  This acts to lower the 
surface energy much like what happens in the random field Ising model(31). Wetting for a random system gives a surface tension that depends on the 
radius of the drop.  This $r$ dependent energy yields $s_c^{-1}$ scaling when the thermodynamic critical exponents for the random first order transition are used.  We now reprise this argument based on a similar one for the random field Ising magnet(31) in Figure 1 (b).

The wetting argument leads to 
a differential renormalization group equation for $\sigma(r)$,
\begin{equation}\label{eq:dfst}
\sigma ^{1/3} d\sigma =-(4^{-1/3}-4^{-4/3})(T\sqrt{k_B\Delta \tilde{c_p}})^{4/3} r^{-5/3}dr,
\end{equation}
where $\Delta \tilde{c_p}$ is the heat capacity jump per unit volume.
This renormalization group equation is 
integrated outward from $r_0$ where the short range value is set by the naive density 
functional theory without wetting discussed earlier, $\sigma _0$.  Between T$_K$ and T$_A$, $\sigma (r)$ 
vanishes at 
large distance and is only finite below T$_K$.  Using this boundary condition, the solution of the renormalization equation for $\sigma (r)$ at 
T$_K$ is then
\begin{equation}\label{eq:sftn2} 
\sigma(r)=\sigma _0(\frac{r_0}{r})^{1/2} .
\end{equation}
When this is substituted into the expression for $F(r)$, one 
finds that the maximum gives a barrier, $\Delta F^{\ddagger}$ which now varies inversely to the first 
power 
of the configurational entropy density; i.e., the Vogel-Fulcher scaling.  We find a simple expression for the activation barrier: 
\begin{equation}\label{eq:fnbr}
\Delta F^{\ddagger}=\frac{3\pi\sigma^2_0r_0}{ Ts_c}=\frac{3\pi\sigma^2_0r_0}{T\Delta \tilde{c_p}} \frac{T_K}{T-
T_K}=k_BTD\frac{T_K}{T-T_K}.
\end{equation}
The coefficient $D$ in this expression has been called the liquid's fragility, which has the expression
\begin{equation}\label{eq:dval}
D=\frac{27}{16}\pi \frac{nk_B}{\Delta \tilde{c_p}}\ln ^2\frac{\alpha _Lr_0^2}{\pi e}.
\end{equation}
Based on the Lindemann ratio universality, the root mean square displacement, $\alpha _L ^{-1/2}$ is taken as $0.1 r_0$, the hard sphere value, so that $D$ can be expressed in terms of the heat capacity jump per mole, $\Delta c_p$, \begin{equation}\label{eq:dval2}
D=32R/ \Delta c_p,
\end{equation}
where $R=8.31$ J mole$^{-1}$ K$^{-1}$.
The value of $D$ depends on the heat capacity jump per mole which varies greatly from substance to substance and is far from being universal.  In Figure 2 we plot the $D$ 
predicted from this theory versus the inverse of the configurational heat 
capacity for several glass forming liquids.  The straight line is given by Equation (8).  Superimposed on the plot are 
the experimental values of the $D$.  The agreement is excellent.  
We see that the magnitude of the activation barriers for rearrangement of the viscous liquid depends on the difference in temperature from T$_K$, on universal microscopic parameters 
connected with the Lindemann ratio and on the excess heat capacity 
connected with configurational excitations.

A hallmark of the random first order transition theory of glass dynamics is the dynamic heterogeneity required to explain the growing barriers upon cooling.  After combining Equation 3 and 5 with our expression of $\sigma _0$, a little algebra shows the characteristic size of a rearranging region is
$\frac{\xi}{r_0}=2(\frac{2}{3\pi \ln \frac{\alpha_Lr_0^2}{\pi e}})^{2/3}(\frac{DT_K}{T-
T_K})^{2/3}$.  This can be expressed as a universal function of the relaxation time $\frac{\xi}{r_0}=2(\frac{2}{3\pi \ln \frac{\alpha_Lr_0^2}{\pi e}})^{2/3}(\ln \frac{\tau}{\tau_0})^{2/3}$, since $\tau =\tau _0exp(\frac{DT_k}{T-T_k})$ according to the Vogel-Fulcher law.  This is plotted in Figure 3.

The kinetic laboratory glass transition occurs when molecular slowing gives relaxations in the hours range, i.e.,  $\frac{\tau}{\tau_0}=10^{17}$.  Thus at T$_g$, $(\frac{\xi}{r_0})\approx 4.5$, a rather modest size.  We also note the universality of $\sigma_0$ suggests that $s_c$ is nearly the same for all substances at the laboratory glass transition.  Roughly 90 molecules are involved in a rearranging unit according to the random first order transition theory at the conventionally defined glass transition temperature.  The rearranging unit according to the Adam-Gibbs argument is a region just capable of having two states, therefore $(\frac{\xi _{AG}}{r_0})=(\frac{R \ln 2}{s_c})^{1/3}$.  $\xi _{AG}$ grows slowly as T$_K$ is approached in contrast to the random first order transition theory.  The Adam-Gibbs argument gives rearranging units with at most 10 molecules near T$_g$ for the most fragile substances.  It is clearly very ambiguous to have such small ``cooperative" units.  Recent observations of structural heterogeneities are inconsistent with units of the small size predicted by Adam-Gibbs, but are in harmony with the estimates of the random first order transition entropic droplet picture(32).  A single size does not characterize the viscous liquid completely.  The random first order transition entropic droplet picture actually leads to a ``mosaic" structure of the supercooled liquid(33) with cooperative regions only somewhat larger than the critical droplet size $\xi$.  These regions fluctuate in size and therefore have different flipping rates because of the configurational entropy fluctuations whose magnitude depends on the configurational heat capacity density jump $\Delta \tilde{c_p}$ and the volume of the rearranging region, $\Delta S_c=\sqrt{k_B\Delta \tilde{c_p} \xi ^3}$.  At $T_g$ both strong and fragile liquids have roughly the same absolute scale for their mosaic structures, i.e., $\xi$ is nearly universal at the laboratory glass temperature. It follows that the range of activation barriers is smaller for strong than for fragile liquids because of their smaller $\Delta \tilde{c_p}$.  This is in accord with the observed correlation between growing nonexponentiality of relaxation with growing fragility(34).

We conclude that the wide variety of kinetic behavior seen in liquids reflects quantitative rather than qualitative differences in their energy landscape.  Furthermore the random first order transition approach coupled with microscopic considerations about the stability of aperiodic structures can account semi-quantitatively for these differences.

\newpage
{\bf Acknowledgment.}  P.G.W. gratefully acknowledges stimulating discussions with Shoji Takada.  This work was supported by NSF grant CHE-9530680.


\begin{thebibliography}{10}

\bibitem{EAN96}
Ediger, M.~D., Angell, C.~A. \& Nagel, S.~R. (1996)
\newblock {\em J.\ Phys.\ Chem.} {\bf 100}, 13200--13212.

\bibitem{KW87a}
Kirkpatrick, T.~R. \& Wolynes, P.~G. (1987)
\newblock {\em Phys.\ Rev.\ A} {\bf35}, 3072--3080.

\bibitem{KT87}
Kirkpatrick, T.~R. \& Thirumalai, D. (1987)
\newblock {\em Phys.\ Rev.\ Lett.} {\bf 58}, 2091--2094.

\bibitem{KW87b}
Kirkpatrick, T.~R. \& Wolynes, P.~G. (1987)
\newblock {\em Phys.\ Rev.\ B} {\bf 36}, 8552--8564.

\bibitem{KTW89}
Kirkpatrick, T.~R., Thirumalai, D. \& Wolynes, P.~G. (1989)
\newblock {\em Phys.\ Rev.\ A} {\bf 40}, 1045--1054.

\bibitem{CKMP96}
Cugliandolo, L.~F., Kurchan, J., Monasson, R. \& Parisi, G. (1996)
\newblock {\em J.\ Phys.\ A} {\bf 29}, 1347--1358.

\bibitem{MP99}
M\'{e}zard, M.\& Parisi, G. (1999)
\newblock {\em Phys.\ Rev.\ Lett.} {\bf 82}, 74--750.

\bibitem{BGS84}
Bengtzelius, U., G\"{o}tze, W. \& Sj\"{o}lander, A. (1984) 
\newblock {\em J.\ Phys.\ C} {\bf 17}, 5915--5934.

\bibitem{SW84}
Stoessel, J.~P. \& Wolynes, P.~G. (1984)
\newblock {\em J.\ Chem.\ Phys.} {\bf 80}, 4502--4512.

\bibitem{SSW85}
Singh, Y., Stoessel, J.~P. \& Wolynes, P.~G. (1985)
\newblock {\em Phys.\ Rev.\ Lett.} {\bf 54}, 1059--1062.

\bibitem{GKS85}
Gross, D.~J., Kanter, I. \& Sompolinsky, H. (1985) 
\newblock {\em Phys.\ Rev.\ Lett.} {\bf 55}, 304--307.

\bibitem{Gotze91}
G\"{o}tze, W. (1991)
\newblock in {\em Liquids, Freezing and the Glass Transition,}
\newblock ed. Hansma, J. P., Levesque, D. \& Zinn-Justin, J. (North-Holland, Amsterdam), pp. 287--504 

\bibitem{CS72}
Champeney, D.~C. \& Sedgwich, D.~F. (1972)
\newblock {\em J.\ Phys.\ C} {\bf 5}, 1903--1913.

\bibitem{Mezei91}
Mezei, F. (1991)
\newblock in {\em Liquids, Freezing and the Glass Transition,}
\newblock ed. Hansma, J. P., Levesque, D. \& Zinn-Justin, J. (North-Holland, Amsterdam,), pp. 629-688. 

\bibitem{DV99}
Dasgupta, C. \& Valls, O.~T. (1999)
\newblock {\em Phys.\ Rev.\ E} {\bf 59}, 3123--3134.

\bibitem{MPV87}
M\'{e}zard, M., Parisi, G. \& Virasoro, M.~A. (1987)
\newblock {\em Spin Glass Theory and Beyond}
\newblock (World Scientific, Singapore).

\bibitem{Gardner85}
Gardner, E. (1985)
\newblock {\em Nucl.\ Phys.\ B} {\bf 257}, 747--765.

\bibitem{Kauz48}
Kauzmann, W. (1943)
\newblock {\em Chem.\ Rev.} {\bf 34}, 219--256.

\bibitem{BM94}
Bouchaud, J.~P. \& M\'{e}zard, M. (1994)
\newblock {\em J.\ Phys.\ I(France)} {\bf 4}, 1109--1114.

\bibitem{FH95}
Franz, S. \& Hertz, J. (1995)
\newblock {\em Phys.\ Rev.\ Lett.} {\bf 74}, 2114--2117.

\bibitem{CIS95}
Chandra, P., Ioffe, L.~B. \& Sherrington D. (1995) 
\newblock {\em Phys.\ Rev.\ Lett.} {\bf 75}, 713--716.

\bibitem{Parisi97}
Parisi, G. (1997)
\newblock {\em J.\ Phys.\ A} {\bf 30}, 8523--8529.

\bibitem{SDS98}
Sastry, S., Debenedetti, P.~G. \& Stillinger, F.~H. (1998)
\newblock {\em Nature} {\bf 393}, 554--557.

\bibitem{BDBG99}
Bennemann, C., Donati, C., Baschnagel J. \& Glotzer, S.~C. (1999)
\newblock {\em Nature} {\bf 399}, 246--249.

\bibitem{AG65}
Adam, G. \& Gibbs, J.~H. (1965)
\newblock {\em J.\ Chem.\ Phys.} {\bf 43}, 139--146.

\bibitem{Oxtoby91}
Oxtoby, D.~W. (1991)
\newblock in {\em Liquids, Freezing and the Glass Transition,}
\newblock ed. Hansma, J. P., Levesque, D. \& Zinn-Justin, J. (North-Holland, Amsterdam), pp. 145-191. 

\bibitem{TW97}
Takada, S. \& Wolynes, P.~G. (1997)
\newblock {\em J.\ Chem.\ Phys.} {\bf 107}, 9585--9598.

\bibitem{Parisi95}
Parisi, G. (1995)
\newblock in {\em Proceedings of the Symposium, The Oskar Klein Century,} 
\newblock (World Scientific, Singapore), pp. 60-71.

\bibitem{Angell84}
Angell, C. (1984)
\newblock in {\em Relaxations in Complex Systems,}
\newblock  ed. Ngai, K. L. \& Wright, G. B. (National Technical Information Service, U.S. Department of Commerce, Springfield, VA), pp. 3-11.

\bibitem{Nieu98}
Nieuwenhuizen, Th.~M. (1998)
\newblock  {\em Phys.\ Rev.\ Lett.} {\bf 81} 2201--2204.

\bibitem{Villain85}
Villain, J. (1985)
\newblock {\em J.\ Phys.} {\bf 46}, 1843--1852.

\bibitem{TWHFSS98}
Tracht, U., Wilhelm, M., Heuer, A., Feng, H., Schmidt-Rohr, K. \& Spiess, H.~W. (1998)
\newblock {\em Phys.\ Rev.\ Lett.} {\bf 81}, 2727--2730.

\bibitem{Wolynes89}
Wolynes, P.~G. (1989)
\newblock in {\em Proceedings International Symposium on Frontiers in Science,}
\newblock  ed. Frauenfelder, H., Chan, S. \& DeBrunner, P.~G. (Am. Inst. Physics), pp. 38-65.

\bibitem{BNAP93}
B\"{o}hmer, B., Ngai, K.~L., Angell, C.~A. \& Plazek, D.~J. (1993) 
\newblock {\em J.\ Chem.\ Phys.} {\bf 99}, 4201--4209.


\end{thebibliography}

\newpage
{\bf Figure 1 a}, Free energy as a function of order parameter $\alpha$.  Right below $T_A$, a second minimum emerges around $\alpha \approx \alpha_L$, which corresponds to a glassy state.  The free energy difference between the liquid and glass state is $TS_c(T)$, which approaches zero at the Kauzmann temperature $T_K$.  {\bf b}, An illustration of a liquid-like (multiconfiguration) droplet inside a glassy region corresponding to a single mean field minimum free energy configuration.  The interface is wetted by suitable configurations to lower the surface energy. One considers an inhomogeneous situation with single minimum given by the density functional theory 
abutting another minimum as in a naive droplet solution with a radius of curvature 
$r$.  Upon this surface, one erects a smaller droplet of one of the other 
solutions of the density functional theory as shown in the figure.  The free 
energy of interpolating this wetting phase is given by
$\delta F=\sigma (r) r^{d-1}(\frac{\zeta }{r})^2-\delta s_cr^{d/2}(\frac{\zeta }{r})^{1/2}$.
This additional free energy cost depends on the surface tension at the scale 
$r$,  $\sigma (r)$ and on the fluctuations in driving force for forming this 
smaller wetting droplet.  In the Ising model, the fluctuations in driving force 
for these droplets arise from the random part of the magnetic field.  For a random first order 
transition, the field fluctuation's role in the disordered magnet is played by the fluctuations of 
configurational entropy density.  The magnitude of these fluctuations 
should be given by the usual Landau expression $\Delta S_c^2=k_B\Delta C_p$ where $\Delta C_p$ is the configurational heat capacity of a region.  The contribution to the free energy from the interpolating wetting droplet yields a change with size of the surface tension at size $r$, $d\sigma$ (Equation 4).

\vspace{2cm}

{\bf Figure 2} The fragility parameter $D$ as a function of the inverse heat capacity jump per mole.  The glass formers chosen are those shown in Angell's review article ({\bf a}. Angell, C.~A. Formation of glasses from liquids and biopolymers. {\em Science} {\bf 267}, 1924--1935 (1995)).  The solid line in the graph is calculated with random first order transition model (Equation (7) and (8)) based on the universality of the Lindemann ratio and the points are from experiments.  Data for fragility parameter $D$ are found in {\bf a}, {\bf b}. Korus, J., Hempel, E., Beiner, M., Kahle, S. \& Donth, E. Temperature dependence of $\alpha$ glass transition cooperativity. {\em Acta Polymer} {\bf 48}, 369--378 (1997), {\bf c}. Richert, R. \& Angell, C.~A. Dynamics of glass forming liquids V. On the link between molecular dynamics and configurational entropy. {\em J.\ Chem.\ Phys.} {\bf 109}, 9016--9026 (1998) and references therein; for specific heat jump are found in {\bf d}. Angell, C.~A. \& Smith, D.~L. Test of the entropy basis of the Vogel-Tammann-Fulcher equation: Dielectric relaxation of polyalcohols near T$_g$. {\em J.\ Phys.\ Chem.} {\bf 86}, 3845--3852 (1982), {\bf e}. Angell, C.~A. \& Torell, L.~M. Short time structural relaxation process in liquids: Comparison of
experimental and computer simulation glass transitions on picosecond time scales. {\em J.\ Chem.\ Phys.} {\bf 78}, 937--945 (1983), and {\bf f}. Torell, L.~M., Ziegler, D.~C. \& Angell, C.~A. Short time relaxation processes in liquids from viscosity and light scattering studies in molten KCl $\cdot$ 2BiCl$_3$. {\em J.\ Chem.\ Phys.} {\bf 81}, 5053--5058 (1984).  The heat capacity jump is given as per mole ``beads" ({\bf d} and {\bf g}. Wunderlich,  B. Study of the change in specific heat of monomeric and polymeric glasses during the glass transition. {\em J.\ Phys.\ Chem.} {\bf 64}, 1052--1056 (1960)
 or ``mobile units" ({\bf h}. Schulz, M. Energy landscape, minimum points, and non-Arrhenius behavior of supercooled liquids. {\em Phys.\ Rev.\ B} {\bf 57}, 11319--11333 (1998)).  Generally speaking, ``beads" are ``rearrangeble elements in a relaxing liquid" ({\bf d}).  We have used the bead count of previous workers ({\bf g}).  The number of beads for GeO$_2$ and ZnCl$_2$ is 3 ({\bf d}), 6 for glycerol ({\bf d}), 10 for KCl$\cdot$ 2BiCl$_3$ ({\bf f}), 12 for 3KNO$_3$ $\cdot$2Ca(NO$_3$)$_2$ ({\bf e}), 2 for m-fluorotoluene ({\bf g} and {\bf I}. Chang, S.~S. \& Bestul, A.~B. Heat capacity and thermodynamic properties of o-terphenyl crytal, glass, and liquid. {\em J.\ Chem.\ Phys.} {\bf 56}, 503--516 (1972)), and 3 for o-terphenyl ({\bf i}).  The ``bead" count is a crude way of accounting for internal flexibility of the molecules (since the free energy functional is essentially that for a monatomic fluid).  To illustrate the robustness of the correlation we indicate also the values of o-terphenyl as shown by a star ($\ast $) if its internal flexibility is ignored and it is assigned a bead count of one.

\vspace{2cm}

{\bf Figure 3} The correlation length $\xi$ (in the unit of lattice spacing $r_0$) is shown as a function of relaxation time.  The solid line is that predicted by random first order transition theory while the dashed line is the result of Adam-Gibbs theory(25) assuming $\Delta c_p=51.8$ J mole$^{-1}$ K$^{-1}$, the value for PVAC (polyvinyl acetate)(25).  The Adam-Gibbs result weakly depends on fragility.  The point (and its error bar) gives the only ``direct" measurement by Spiess {\it et al}. on PVAC(32).  Results of many ``indirect" measurements on different glass formers around the glass transition temperature also fall in the range of the Spiess experiment(32).  

\newpage
\begin{figure}[t]
\hspace{16cm}
\psfig{file=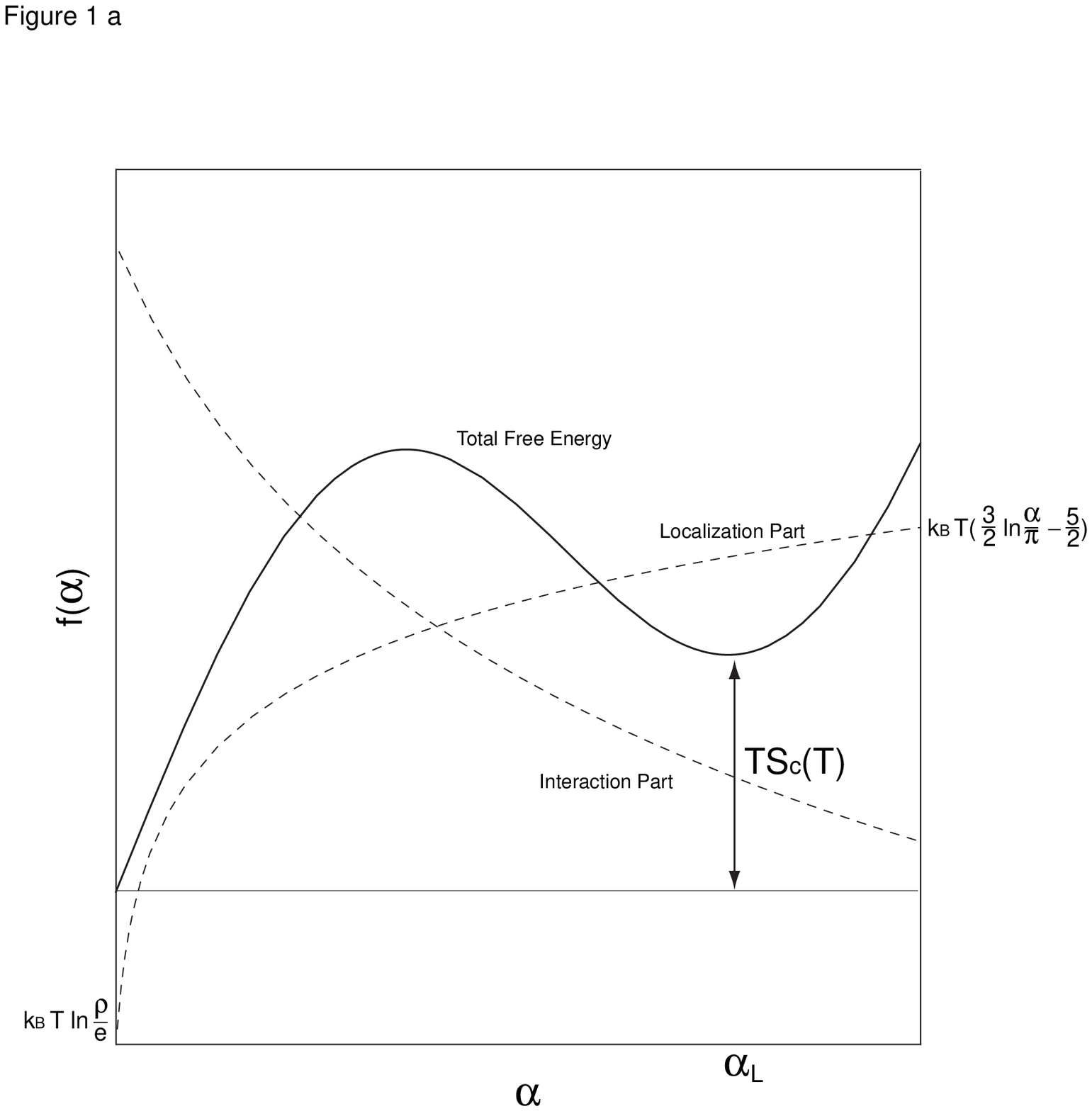}
\end{figure}

\newpage
\begin{figure}[htb]
\hspace{.2cm}
\psfig{file=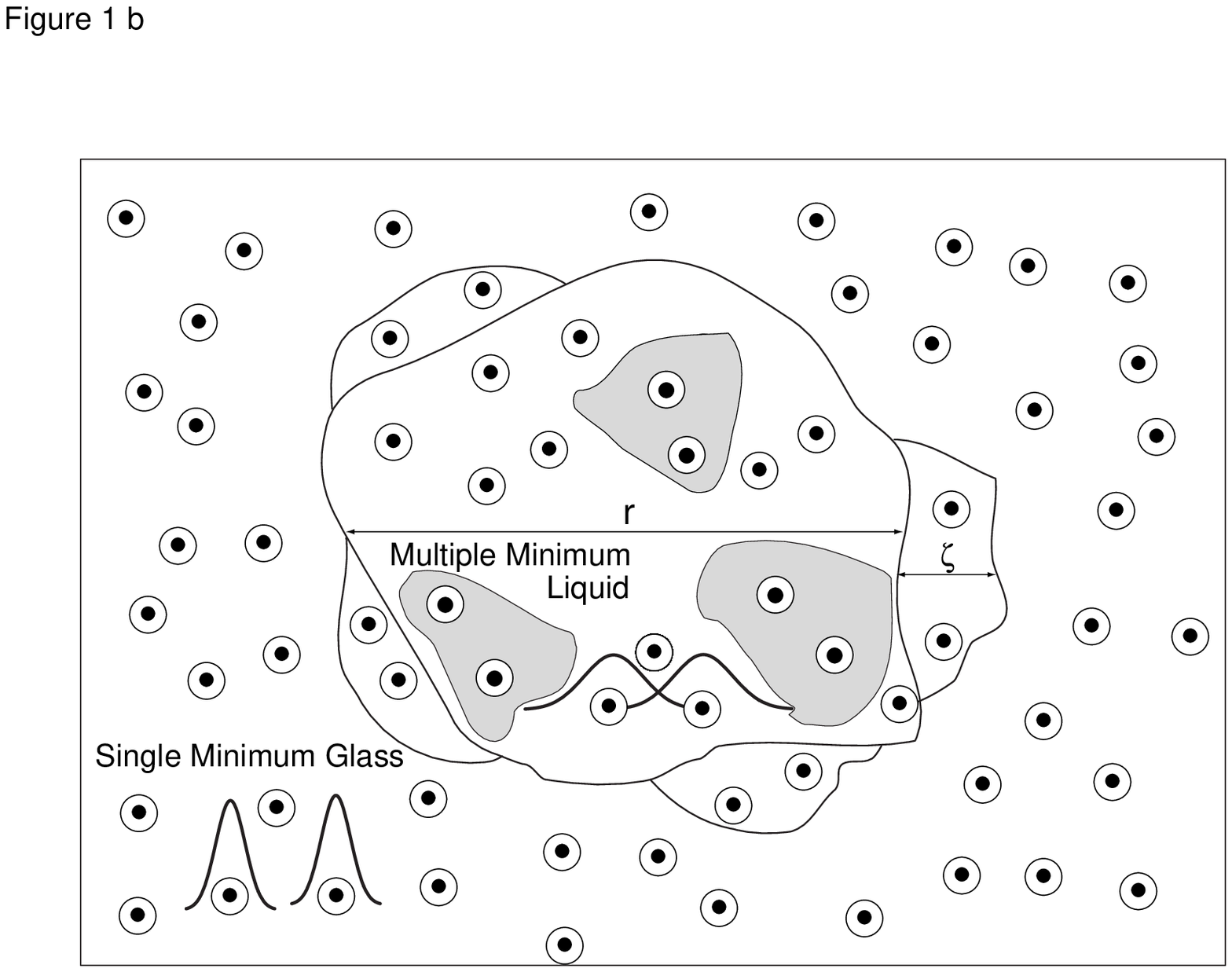}
\end{figure}

\newpage
\begin{figure}[htb]
\hspace{.2cm}
\psfig{file=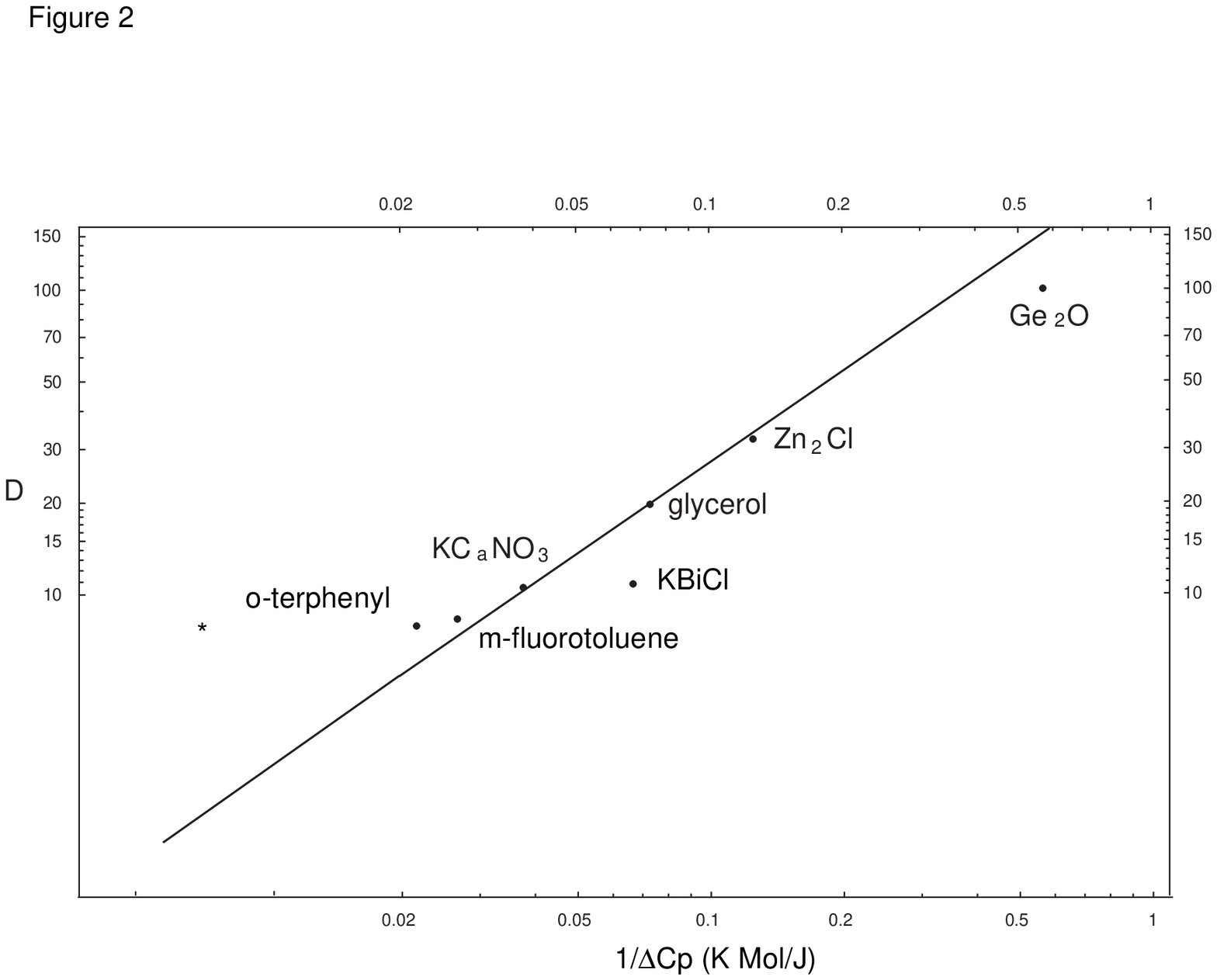}
\end{figure}

\newpage
\begin{figure}[htb]
\hspace{.2cm}
\psfig{file=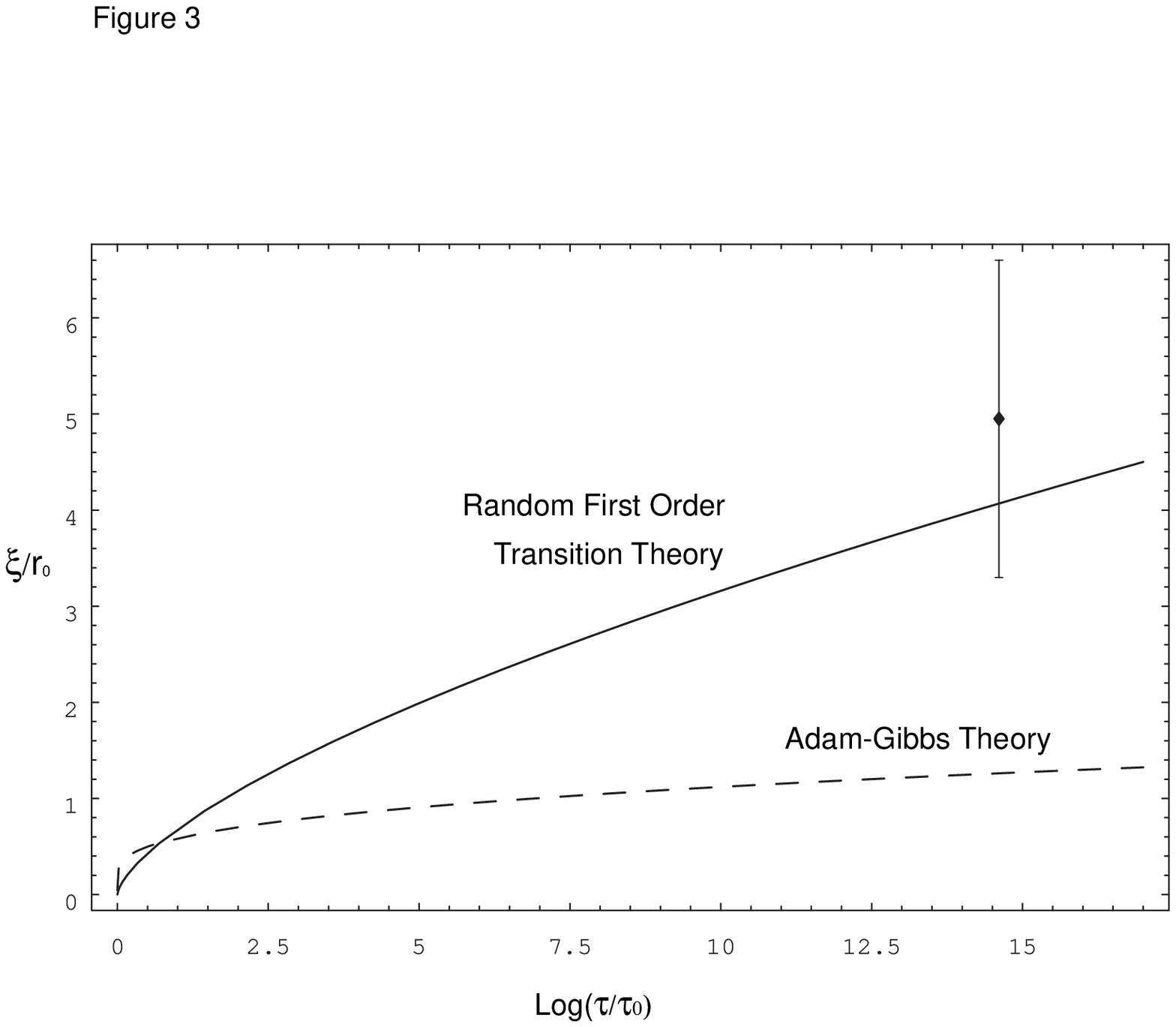}
\end{figure}
 
\end{document}